\documentclass[twocolumn,prb,showpacs,amsmath]{revtex4}
\usepackage{epsfig}

\def\be{\begin{equation}}
\def\ee{\end{equation}}
\def\e#1{\label{#1}\end{equation}}
\def\bea{\begin{eqnarray}}
\def\eea{\end{eqnarray}}
\def\ea#1{\label{#1}\end{eqnarray}}
\def\rqn#1{(\ref{#1})}
\def\bes#1{\begin{subequations}\label{#1}}
\def\ese{\end{subequations}}

\begin{document}

\title{Analysis of Bell inequality violation in superconducting
phase qubits}
\author{Abraham G. Kofman}
\author{Alexander N. Korotkov}
\affiliation{Department of Electrical Engineering, University of
California, Riverside, California 92521}
\date{\today}

\begin{abstract}
We analyze conditions for violation of the Bell inequality in the
Clauser-Horne-Shimony-Holt form, focusing on the Josephson phase
qubits. We start the analysis with maximum violation in the ideal
case, and then take into account the effects of the local
measurement errors and decoherence. A special attention is paid to
configurations of the qubit measurement directions in the pseudospin
space lying within either horizontal or vertical planes; these
configurations are optimal in certain cases. Besides local
measurement errors and decoherence, we also discuss the effect of
measurement crosstalk, which affects both the classical inequality
and the quantum result. In particular, we propose a version of the
Bell inequality which is insensitive to the crosstalk.
 \end{abstract}
 \maketitle

\section{Introduction}

In 1935 Einstein, Podolsky, and Rosen (EPR) have shown in the
classical paper \cite{EPR} that quantum mechanics contradicts the
natural assumption (the ``local realism'') that a measurement of one
of two spatially separated objects does not affect the other one.
 This ``spooky action at a distance'' -- known as an
entanglement -- is now recognized as a major resource in the field
of quantum information and quantum computing. \cite{nie00}
 The paradox has led EPR to conclude that quantum mechanics is an incomplete description
of physical reality, thus implying that some local hidden variables
are needed.

 The EPR paradox remained at the level of semi-philosophical discussions until 1964, when
John Bell contributed an inequality for results of a
spin-correlation experiment, \cite{Bell} which should hold for any
theory involving local hidden variables, but is violated by quantum
mechanics.
 Inspired by Bell's idea, in 1969, Clauser, Horne, Shimony, and Holt
\cite{CHSH} (CHSH) proposed a version of the Bell inequality (BI,
the generic name for a family of inequalities), which made the
experimental testing of local hidden-variable theories possible. The
main advantage of the CHSH inequality in comparison with the
original BI is that it does not rely on an experimentally
unrealistic assumption of a perfect anticorrelation between the
measurement results when two spin-1/2 particles (in the spin-0
state) are measured along the same direction.
 Many interesting experiments
\cite{Freedman,asp81,Ou,Rowe,moe04,Hasegawa03} have been done since
then.
 The results of these experiments clearly show a violation of the
Bell inequalities, in accordance with quantum mechanical
predictions.

The BI violation has been mostly demonstrated in the experiments
with photons;\cite{Freedman,asp81,Ou} it has been also shown in the
experiments with ions in traps \cite{Rowe} and with an atom-photon
system; \cite{moe04} a Bell-like inequality violation has been also
demonstrated in an experiment with single neutrons.\cite{Hasegawa03}
Experimental violation of the BI in solid-state qubits would be an
important step towards practical quantum information processing by
solid-state devices.\cite{Pashkin,mcd05,Mooij,Schoelkopf,Marcus}
Experiments on observation of the BI violation in Josephson phase
qubits are currently underway. \cite{ans06,ans07} Theoretical study
related to the BI violation in solid-state systems has also
attracted significant attention in recent years. \cite
{Nori-05,Nori-06,Ionicioiu,sam03,Beenakker-03,Lebedev-04,Samuelsson-05,Trauzettel-06}

In this paper we discuss the Bell inequality (in the CHSH form) for
solid-state systems, focusing on experiments with superconducting
phase qubits.
 We study effects of various factors detrimental for observation of
the BI violation, including local measurement errors, decoherence,
and interaction between qubits (crosstalk), and analyze optimal
conditions in presence of these nonidealities.

The paper is organized as follows.
 In Sec.\ \ref{II} we review the CHSH type of the BI, and
also discuss tomography-type measurements using qubit rotations.
 In Sec.\ \ref{III} we consider the ideal case and describe all
situations for which the BI is violated maximally.
 Sections \ref{IV}-\ref{VII} are devoted to the effects of various
nonidealities on the observation of the BI violation.
 In Sec.\ \ref{IV} we discuss the effect of local
measurement errors, using a more general error model than in the
previous approaches. \cite{ebe93,kwi93,zuk93,lar01,cab07} Analytical
results for maximally entangled states and numerical results for
general two-qubit states are presented.
 In Sec.\ \ref{VI} we consider the effect of local decoherence of
the qubits.
 In Sec.\ \ref{VII} we discuss measurement crosstalk, which
affects both the BI (since crosstalk is a classical mechanism of
communication between qubits) and the quantum result. We also
propose a version of the BI, which is not affected by the crosstalk.
 Section \ref{VIII} provides concluding remarks.

\section{Preliminaries}
 \label{II}

\subsection{CHSH inequality}

We begin with a brief review of the CHSH inequality,
\cite{CHSH,cla78,asp02} a type of the Bell inequality usually used
in experiments.
 Let us consider a pair of two-level systems (qubits) $a$ and $b$.
Assuming that a realistic (classical) theory based on local
observables \cite{Bell} holds and there is no communication between
the qubits (i.e., no crosstalk), the two-qubit measurement results
should satisfy the CHSH inequality \cite{CHSH,bel71}
 \be
-2\le S\le2,
 \e{3}
where
 \be
S=E(\vec{a},\vec{b})-E(\vec{a},\vec{b}')+E(\vec{a}',\vec{b})+
E(\vec{a}',\vec{b}').
 \e{4}
Here $\vec{a}$ and $\vec{a}'$ ($\vec{b}$ and $\vec{b}'$) are the
unit radius-vectors on the Bloch sphere along the measurement axes
for qubit $a$ ($b$) and $E(\vec{a},\vec{b})$ is the correlator of
the measurement results:
 \be
E(\vec{a},\vec{b})=p_{++}(\vec{a},\vec{b})+p_{--}(\vec{a},\vec{b})-
p_{+-}(\vec{a},\vec{b})-p_{-+}(\vec{a},\vec{b}),
 \e{5}
where $p_{ij}(\vec{a},\vec{b})\ (i,j=\pm)$ is the joint probability
of measurement results $i$ and $j$ for qubits $a$ and $b$,
respectively.

The sum of the probabilities in Eq.\ \rqn{5} equals one.
 This can be used to recast Eq.\ \rqn{4} in the form
 \be
S=4T+2,
 \e{22}
where
 \be
T=p(\vec{a},\vec{b})-p(\vec{a},\vec{b}')+p(\vec{a}',\vec{b})+
p(\vec{a}',\vec{b}')-p_a(\vec{a}')-p_b(\vec{b}).
 \e{7}
Here $p(\vec{a},\vec{b})=p_{++}(\vec{a},\vec{b} )$, whereas
$p_a(\vec{a}')=p_{++}(\vec{a}',\vec{b})+p_{+-}(\vec{a}',\vec{b})$
[or $p_b(\vec{b})=p_{++}(\vec{a},\vec{b})+p_{-+}(\vec{a},\vec{b})$]
is the probability of the measurement result ``+'' for qubit $a$ (or
$b)$ irrespective of the measurement result for the other qubit [in
the classical theory in absence of communication between qubits,
$p_a(\vec{a}')$  is obviously independent of the direction $\vec{b}$
and even independent of the very fact of the qubit $b$ measurement;
similarly for $p_b(\vec{b})$].
 Thus, instead of the probabilities $p_{ij}$, one can equivalently
use the probabilities $p$, $p_a$, and $p_b$, and the inequality
\rqn{3} can be recast in an equivalent form \cite{CHSH}
 \be
-1\le T\le0.
 \e{6}

Notice that both inequalities \rqn{3} and \rqn{6} can have somewhat
different meanings in different physical situations. In particular,
the results ``$+$'' and ``$-$'' may correspond to the presence or
absence of the detector ``click'' (so called ``one-channel
measurement'' \cite{CHSH,cla74}); in this case a low-efficiency
detector significantly increases the chance of the result ``$-$''.
Another possibility is the so-called ``two-channel measurement'', in
which the qubit states``$+$'' and ``$-$'' are supposed to produce
clicks in different detectors; \cite{bel71} in this case inefficient
detection leads to three possible results: ``$+$'', ``$-$'', and
``no result''. Significant inefficiency of the optical detectors
leads to the so-called ``detector loophole,'' \cite{pea70,gar87}
which arises because effectively not the whole ensemble of the qubit
pairs is being measured. This problem is often discussed in terms of
contrasting the Clauser-Horne \cite{cla74} (CH) and CHSH
interpretations of the inequalities, which differ by considering
either the whole ensemble or a subensemble of the qubit pairs. It is
important to mention that in the case of the Josephson phase qubits
(which formally belongs to the class of one-channel measurements)
the whole ensemble of qubit pairs is being measured, and therefore
there is no detector loophole (if one avoids \cite{ans07}
corrections for measurement errors), as well as there is no
difference between CHSH and CH interpretations.

\subsection{Tomographic measurements}

In some cases, as for Josephson phase qubits, the measurement
(detector) axis cannot be physically rotated.
 However, instead of the detector rotation, one can rotate the qubit state.
\cite{negative-result,ste06}
 Let us show the equivalence of the two methods explicitly, using the
example of the phase qubit and assuming ideal (orthodox)
measurement.

The Hamiltonian of the phase qubit in a microwave field in the
subspace of the two lowest states in the qubit potential well
$|\psi_0\rangle$ and $|\psi_1\rangle$ is
 \bea
&H_q=&(\hbar\omega_q/2)(|\psi_1\rangle\langle\psi_1|
-|\psi_0\rangle\langle\psi_0|)\nonumber\\
 &&+\hbar\Omega(t)\sin(\omega t+\phi)
(|\psi_0\rangle\langle\psi_1|+|\psi_1\rangle\langle\psi_0|),
 \ea{9}
where $\hbar$ is the Planck constant, $\omega_q$ is the qubit
resonance frequency, $\omega$ is the microwave frequency,
$\Omega(t)=d_{10}E(t)/\hbar$ is the time-dependent Rabi frequency,
$d_{10}$ is the dipole-moment matrix element, and $E(t)$ is the
electric-strength amplitude of the microwave field.
 Transforming to the qubit basis (the rotating frame)
$|0\rangle=|\psi_0\rangle$, $|1\rangle=e^{i\omega t}|\psi_1\rangle$,
neglecting fast oscillating terms in the Hamiltonian (the
rotating-wave approximation) and assuming the resonance condition
$\omega=\omega_q$, we arrive at the Hamiltonian
 \be
H(t)=\frac{\hbar\Omega(t)}{2}\,\vec{n}\cdot\vec{\sigma},
 \e{10}
where the unit vector $\vec{n}=(\sin\phi,-\cos\phi,0)$ lies in the
$xy$ plane making the angle $\phi-\pi/2$ with the $x$ axis, whereas
$\vec{\sigma}=(\sigma_x,\sigma_y,\sigma_z)$ is the vector of the
Pauli matrices. \cite{lan77}
 Here and below we associate state $|1\rangle$ ($|0\rangle$) with
the measurement result ``$+$'' (``$-$'') and with the eigenvalue $1$
($-1$) of $\sigma_z$, so that
$\sigma_z=|1\rangle\langle1|-|0\rangle\langle0|$.

 As follows from Eq.\ \rqn{10}, the microwave pulse rotates the qubit
state such that the initial density matrix $\rho_q$ becomes
$\tilde{\rho}_q=U_R\rho_qU_R^\dagger$, where
 \bea
&U_R&=e^{-i\theta\vec{n}\cdot\vec{\sigma}/2}
=\cos(\theta/2)-i\vec{n}\cdot\vec{\sigma}\sin(\theta/2)\nonumber\\
&&=\left(\begin{array}{cc}
\cos(\theta/2)&e^{-i\phi}\sin(\theta/2)\\
-e^{i\phi}\sin(\theta/2)&\cos(\theta/2)
\end{array}\right).
 \ea{11}
Here $\theta=\int_{t_1}^{t_2}dt\,\Omega(t)$, where $t_1$ and $t_2$
are the pulse starting and ending time moments.

The probability of the qubit to be found in state $|i\rangle$ is
 \be
p_i={\rm Tr}(|i\rangle\langle i|\tilde{\rho}_q)={\rm Tr}(P_i\rho_q)\
\ (i=0,1).
 \e{12}
Here $P_i$ is the projection operator
$P_i=U_R^\dagger|i\rangle\langle i|U_R$, i.e.,
 \bea
&&P_1(\vec{a})=\frac{1}{2}\left(\begin{array}{cc}
1+\cos\theta&e^{-i\phi}\sin\theta\\
e^{i\phi}\sin\theta&1-\cos\theta
\end{array}\right)=\frac{1}{2}(I+\vec{a}\cdot\vec{\sigma}),
\nonumber\\
&&P_0(\vec{a})=I-P_1(\vec{a})=
\frac{1}{2}(I-\vec{a}\cdot\vec{\sigma}),
 \ea{29}
where $I$ is the identity matrix and $\vec{a}$ is the unit vector
 \be
\vec{a}=(\cos\phi\sin\theta,\sin\phi\sin\theta,\cos\theta)
 \e{2.6}
defining the measurement axis.

 Equations \rqn{12} and \rqn{29} show explicitly the equivalence
between the qubit and detector rotations.
 Namely, one can interpret $p_1$ ($p_0$) as the probability of
the qubit to be found in the state with the pseudospin parallel
(antiparallel) to the measurement axis $\vec{a}$.

    Notice that the microwave phase $\phi$ is naturally defined
modulo $2\pi$, while the Rabi rotation angle $\theta$ can be always
reduced to a $2\pi$ range (we will assume $-\pi<\theta\le\pi$).
Nevertheless, with this restriction there are still two sets of
angles $(\theta , \phi )$ corresponding to the same measurement
direction $\vec{a}$, because $\vec{a}$ is invariant under the change
 \be
(\theta,\phi)\leftrightarrow(-\theta,\phi+\pi).
 \e{2.7}
It is easy to make a one-to-one correspondence between the
measurement direction $\vec{a}$ and angles $(\theta , \phi )$ by
limiting either $\theta$ or $\phi$ to a $\pi$-range (instead of
$2\pi$). However, we prefer not to do that because it is convenient
and natural physically to have a $2\pi$ range for one angle when the
other angle is fixed. So, as follows from Eqs.\ \rqn{2.6} and
\rqn{2.7}, the polar (zenith) and azimuth spherical coordinates of
$\vec{a}$ are equal to, respectively, $\theta$ and $\phi$ when
$\theta\ge0$, and $-\theta$ and $\phi+\pi$ when $\theta<0$.

The joint probability of the two-qubit measurement can be written as
 \be
p_{ij}(\vec{a},\vec{b}) ={\rm Tr}[P_i^a(\vec{a})
P_j^b(\vec{b})\rho],
 \e{13}
where $\rho$ is the two-qubit density matrix, $P_i^a=P_i\otimes I$,
and $P_i^b=I\otimes P_i$.

\section{Maximum BI violation: Ideal case}
\label{III}

The purpose of this paper is to analyze conditions needed to observe
the BI violation in experiment.
 Since it is usually easier to observe an effect when it is
maximal, we start the analysis with the situations where violation
of the BI is maximal.

\subsection{Bell operator and Cirel'son's bounds}

 Equations \rqn{5} and \rqn{13} yield $E(\vec{a},\vec{b})={\rm
Tr}(AB\rho)$, where
 \be
A=P_1^a(\vec{a})-P_0^a(\vec{a})= \vec{a}\cdot\vec{\sigma}_a
 \e{3.24}
and similarly $B=\vec{b}\cdot\vec{\sigma}_b$.
 Here $\vec{\sigma}_a=\vec{\sigma}\otimes I$ and
$\vec{\sigma}_b=I\otimes\vec{\sigma}$, the eigenvalues of $A$ and
$B$ being $\pm1$.
 Correspondingly, as follows from Eq.\ \rqn{4},
 \be
S={\rm Tr}({\cal B}\rho),
 \e{3.5}
where the Bell operator \cite{bra92} ${\cal B}$ is
 \be
{\cal B}=AB-AB'+A'B+A'B'
 \e{2.2}
(here $A'=\vec{a}'\cdot\vec{\sigma}_a$ and
$B'=\vec{b}'\cdot\vec{\sigma}_b$).

    The maximum and minimum values of $S$ (so-called Cirel'son's bounds \cite{cir80})
 \be
S_\pm=\pm2\sqrt{2}
 \e{19}
can be obtained, for example, in the following way. \cite{lan87}
Since the Bell operator ${\cal B}$ is Hermitian, $S_\pm$ are equal
to the maximum and minimum eigenvalues of ${\cal B}$. These
eigenvalues can be found analyzing the eigenvalues of ${\cal B}^2$:
 \be
{\cal B}^2=4+[A,A'][B,B']
=4-4(\vec{a}\times\vec{a}'\cdot\vec{\sigma}_a)\, (\vec{b}
\times\vec{b}'\cdot\vec{\sigma}_b)
 \e{2.3}
(here the vector product is taken before the scalar product). Since
the eigenvalues of the Pauli matrices are equal to $\pm 1$, the
largest eigenvalue of ${\cal B}^2$ is 8, achieved when
$\vec{a}\perp\vec{a}'$ and $\vec{b}\perp\vec{b}'$. In this case the
maximum and minimum eigenvalues of ${\cal B}$ are $\pm \sqrt{8}$,
thus leading to Eq.\ (\ref{19}). (Both values $\pm \sqrt{8}$ are
realized because in this case the other eigenvalue of ${\cal B}^2$
is 0, and the sum of all four eigenvalues of ${\cal B}$ should be
equal to 0 since ${\rm Tr}{\cal B}=0$.)

Consider some useful properties of $S$.
 As follows from Eqs.\ \rqn{3.5} and \rqn{2.2}, the value of $S$ is
invariant under arbitrary local unitary transformations $U_a$ and
$U_b$,
 \bes{3.16}\be
\rho\rightarrow(U_a\otimes U_b)\rho(U_a^\dagger\otimes U_b^\dagger),
 \e{3.16a}
if simultaneously $A\rightarrow U_aAU_a^\dagger,\ A'\rightarrow
U_aA'U_a^\dagger,\ B\rightarrow U_bBU_b^\dagger,\ B'\rightarrow
U_bB'U_b^\dagger$ or, equivalently,
 \bea
\vec{a}\rightarrow R_a\vec{a},\ \ \vec{a}'\rightarrow R_a\vec{a}', \
\ \vec{b}\rightarrow R_b\vec{b},\ \ \vec{b}'\rightarrow R_b\vec{b}',
 \ea{3.16b}
 \ese
where $R_a$ ($R_b$) is the rotation matrix corresponding to $U_a$
($U_b$), so that, e.g.,
$U_a(\vec{a}\cdot\vec{\sigma})U_a^\dagger=(R_a\vec{a})
\cdot\vec{\sigma}$.
    This invariance is an obvious consequence of the equivalence
between the qubit and detector rotations, discussed in the previous
section.
 As a result of the invariance, if some state is known to violate the BI for a
given configuration of the detectors, one can obtain many other
states and the corresponding detector configurations providing the
same BI violation, by using Eqs.\ \rqn{3.16} with all possible local
rotations.

Note also that $\cal B$ inverts the sign if the pair of vectors
$\vec{a},\vec{a}'$ (or $\vec{b},\vec{b}'$) inverts the sign.
 Correspondingly, for a given state
 \be
S\rightarrow -S \,\,\,\, \mbox{if} \,\,\,\,
\vec{a}\rightarrow-\vec{a},\,\vec{a}'\rightarrow-\vec{a}'\
(\mbox{or}\
\vec{b}\rightarrow-\vec{b},\,\vec{b}'\rightarrow-\vec{b}').
 \e{3.25}
As follows from Eq.\ \rqn{3.25}, there is a one-to-one
correspondence between the classes of detector configurations
maximizing and minimizing $S$ for a given state.

\subsection{Optimal detector configurations for maximally
entangled states}
 \label{IIIB}

For any given detector configuration satisfying the condition
 \be
\vec{a}\perp\vec{a}' \quad \mbox{and} \quad \vec{b}\perp\vec{b}',
 \e{2.4}
each of the Cirel'son's bounds \rqn{19} is achieved for a unique
maximally entangled state. \cite{pop92,kar95}
 In contrast, for a given maximally entangled state there can be many
optimal detector configurations giving the maximal BI violation.
 To the best of our knowledge, only the configurations with
the detector axes lying in one plane have been usually considered in
the literature, though generally the detector axes for different
qubits may lie in two different planes.
 Moreover, the BI violation has been studied mainly for one of the Bell
states, \cite{nie00}
 \bea
&&|\Psi_\pm\rangle=(|10\rangle\pm|01\rangle)/\sqrt{2}, \label{28}\\
&&|\Phi_\pm\rangle=(|00\rangle\pm|11\rangle)/\sqrt{2},
 \ea{3.23}
while an arbitrary maximally entangled state can be written as
 \be
|\Psi_{\rm me}\rangle=(|\chi_1^a\chi_1^b\rangle+
|\chi_2^a\chi_2^b\rangle)/\sqrt{2},
 \e{3.18}
where $\{|\chi_1^k\rangle,|\chi_2^k\rangle\}$ is an orthonormal
basis for qubit $k$. Our purpose here is to determine all optimal
detector configurations for any maximally entangled state.

\subsubsection{Singlet state}
\label{IIIB1}

Let us start, assuming that the qubits are in the singlet state
$|\Psi_-\rangle$.
 For this state
$E(\vec{a},\vec{b})=\langle\Psi_-|
(\vec{a}\cdot\vec{\sigma})\otimes(\vec{b}\cdot\vec{\sigma})
|\Psi_-\rangle=-\vec{a}\cdot\vec{b}$, so that [see Eq.\ \rqn{4}]
 \be
S=\vec{a}\cdot(\vec{b}'-\vec{b})-\vec{a}'\cdot(\vec{b}+\vec{b}').
 \e{g1}
Maximizing this formula over $\vec{a}$ and ${\vec{a}'}$, we should
choose $\vec{a}$ to be parallel to $\vec{b}'-\vec{b}$, while
$\vec{a}'$ should be antiparallel to $\vec{b}'+\vec{b}$.
Therefore\cite{pop92a} $\vec{a}\perp\vec{a}'$, since
$\vec{b}'-\vec{b}$ and $\vec{b}+\vec{b}'$ are mutually orthogonal.
Similarly, rewriting $S$ as
$S=\vec{b}'\cdot(\vec{a}+\vec{a}')-\vec{b}\cdot(\vec{a}-\vec{a}')$,
we can show that maximization of $S$ requires
$\vec{b}\perp\vec{b}'$. In this way we easily show that the {\it
necessary and sufficient condition} for reaching the upper bound
$S_+=2\sqrt{2}$ for the singlet state is
 \be
\vec{a}\perp\vec{a}',\ \ \vec{b}=-(\vec{a}+\vec{a}')/\sqrt{2},\ \
\vec{b}'=(\vec{a}'-\vec{a})/\sqrt{2}.
 \e{g2}

Because of the symmetry (\ref{3.25}), the necessary and sufficient
condition for reaching the lower bound $S_-=-2\sqrt{2}$ can be
obtained by the inversion of the detector directions for one of the
qubits:
 \be
\vec{a}\perp\vec{a}',\ \ \vec{b}=(\vec{a}+\vec{a}')/\sqrt{2},\ \
\vec{b}'=(\vec{a}-\vec{a}')/\sqrt{2}.
 \e{g3}

Equations \rqn{g2} and \rqn{g3} show that for the singlet state the
maximum BI violation requires that the detector axes for both qubits
lie in the same plane. However, the orientation of this plane is
arbitrary, since Eqs.\ \rqn{g2} and \rqn{g3} determine only the
angles between the detector axes.

    All configurations maximizing (or minimizing) $S$ for the
singlet state can be obtained from {\em one} maximizing (minimizing)
configuration by all possible rotations of the plane containing the
detector axes. As the initial maximizing case we can choose the most
standard configuration\cite{CHSH,asp02} when all detector axes are
within $xz$ plane,
    \begin{equation}
\phi_a=\phi_a'=\phi_b=\phi_b'=0,
    \label{conf-st-1}\end{equation}
and the polar (zenith) angles of the detector directions $\vec{a}$,
$\vec{a}'$, $\vec{b}$, $\vec{b}'$ are
     \begin{equation}
\theta_a =0, \,\,\, \theta_a'=\pi/2, \,\,\, \theta_b=-3\pi/4, \,\,\,
\theta_b'=-\pi/4 .
    \label{conf-st-2}\end{equation}
Then all detector configurations with $S=2\sqrt{2}$ can be
parametrized by three Euler angles \cite{note4} $\kappa_1$,
$\kappa_2$, and $\kappa_3$ ($0\le\kappa_{1,3}\le2\pi$,
$0\le\kappa_2\le\pi$), which describe an arbitrary rotation of the
configuration (\ref{conf-st-1})--(\ref{conf-st-2}).

    Similarly, all minimizing configurations ($S=-2\sqrt{2}$)  can be obtained from
the standard $xz$ case [Eq.\ (\ref{conf-st-1})] with
     \begin{equation}
\theta_a =0, \,\,\, \theta_a'=\pi/2, \,\,\, \theta_b=\pi/4, \,\,\,
\theta_b'=3\pi/4
    \label{conf-st-3}\end{equation}
by arbitrary rotations of this configuration, characterized by three
Euler angles $\kappa_{1,2,3}$.

\subsubsection{General maximally entangled state}
\label{IIIB2}

Any maximally entangled two-qubit state can be obtained from the
singlet state by a unitary transformation of the basis of one of the
qubits \cite{vol00} (i.e.\ a one-qubit rotation). Therefore, an
arbitrary case corresponding to the bounds $S_{\pm}=\pm 2\sqrt{2}$
can be reduced to the singlet state considered above by a unitary
transformation $U_b$ of the qubit $b$ basis and simultaneous
corresponding rotation of the detector axes $\vec{b}$ and $\vec{b}'$
for the second qubit. Since the transformation $U_b$ can also be
characterized by three Euler angles $\kappa_1^b$, $\kappa_2^b$, and
$\kappa_3^b$, an arbitrary situation with $S=2\sqrt{2}$ can be
characterized by six independent parameters
$(\kappa_1,\kappa_2,\kappa_3,\kappa_1^b,\kappa_2^b,\kappa_3^b)$,
using the standard configuration
(\ref{conf-st-1})--(\ref{conf-st-2}) as a starting point. Similarly,
any situation with $S=-2\sqrt{2}$ is characterized by the same six
parameters, starting with the $xz$ configuration (\ref{conf-st-3}).

   Since these six parameters can describe arbitrary directions
of four measurement axes $(\vec{a},\vec{a}',\vec{b},\vec{b}')$ still
satisfying the conditions $\vec{a}\perp\vec{a}'$ and
$\vec{b}\perp\vec{b}'$, it is obvious that any such four-axes
configuration produces $S=2\sqrt{2}$ for exactly one entangled state
and also produces $S=-2\sqrt{2}$ for another entangled state.
    Notice that the sign of $S$ can obviously be flipped by a
$\pi$-rotation of qubit $a$ (or $b$) around the axis
$\vec{a}\times\vec{a}'$ ($\vec{b}\times\vec{b}'$) instead of the
$\pi$-rotation (\ref{3.25}) of its detector axes. Also notice that
six independent parameters for an optimal configuration can be
alternatively chosen as any parameters characterizing the four
measurement axes, which are pair-wise orthogonal:
$\vec{a}\perp\vec{a}'$ and $\vec{b}\perp\vec{b}'$.

\subsubsection{Odd states}
\label{IIIB3}

   An important special case is the class of ``odd'' maximally
   entangled
 states
  \be
|\Psi\rangle=(|10\rangle+ e^{i\alpha}|01\rangle)/\sqrt{2}\ \
(0\le\alpha<2\pi),
 \e{2.5}
which is of relevance for experiments with Josephson phase qubits.
\cite{mcd05}
 Such states can be obtained (with an accuracy up to an overall phase factor)
from the singlet state $|\Psi_-\rangle$ by unequal rotations of the
two qubits around the $z$ axis.
 Indeed, since $U_z(\varphi)=e^{-i\varphi\sigma_z/2}$
rotates a spin 1/2 around the $z$ axis by angle $\varphi$, we obtain
$[U_z(\alpha_0)\otimes U_z(\alpha_0+\pi-\alpha )]|\Psi_-\rangle
=-ie^{-i\alpha/2}|\Psi\rangle$, where $\alpha_0$ is arbitrary.
 Thus, in view of Eq.\ \rqn{3.16}, the optimal detector
configurations for the odd state \rqn{2.5} can be obtained from
those for $|\Psi_-\rangle$ by rotating the detectors for the qubit
$b$ around the $z$ axis by the angle $\pi-\alpha$ [notice that the
state (\ref{2.5}) reduces to the singlet for $\alpha =\pi$].
 In terms of the parameters $\theta$ and $\phi$, this is equivalent
to the change
 \be
\phi_b\rightarrow\phi_b+\pi-\alpha,\ \
\phi_b'\rightarrow\phi_b'+\pi-\alpha.
 \e{g4}
Thus, for the odd states the class of optimal configurations
maximizing $S$ (as well as the class minimizing $S$) is
characterized by four parameters: $\kappa_1,\ \kappa_2,\ \kappa_3$,
and $\alpha$.

   Now let us focus on the optimal configurations with the detector axes
lying either in a ``vertical'' plane for each qubit (i.e., a plane
containing the $z$ axis) or the ``horizontal'' ($xy$) plane; such
configurations will be important in the study of effects of errors
(Sec.\ \ref{IV}) and decoherence (Sec.\ \ref{VI}).

 To obtain all ``vertical'' cases with the maximum BI violation $S=2\sqrt{2}$,
we start with the standard configuration
(\ref{conf-st-1})--\rqn{conf-st-2} for the singlet, then apply a
rotation in the $xz$ plane by an arbitrary angle $C$ (we can also
apply the mirror reflection), then rotate the resulting
configuration around the $z$ axis by an arbitrary angle $\phi_0$,
\cite{note1} and finally apply the $\alpha$-rotation \rqn{g4}
determined by the phase of the odd state (\ref{2.5}).
 As a result, the optimal measurement directions for the qubits $a$ and $b$
generally lie in different vertical planes,
 \be
\phi_a=\phi_a'=\phi_0,\ \ \phi_b=\phi_b'=\phi_0+\pi-\alpha,
 \e{3.26a}
while the polar angles corresponding to $S=2\sqrt{2}$ are
 \be
(\theta_a,\theta_a',\theta_b,\theta_b')=\pm(0,\pi/2,-3\pi/4,-\pi/4)+C,
 \e{3.4a}
where $\phi_0$ and $C$ are arbitrary angles, while $\alpha$ is
determined by the state (\ref{2.5}). Notice that the signs $\pm$
correspond to the possibility of the mirror reflection, which we did
not have to consider in the previous subsections because it can be
reproduced using 3D rotations, while it is a necessary extra
transformation in the 2D case.

    Similarly, the minimum $S=-2\sqrt{2}$ for the odd state
(\ref{2.5}) is achieved for the vertical configurations within the
planes given by Eq.\ (\ref{3.26a}) for the polar angles
 \be
(\theta_a,\theta_a',\theta_b,\theta_b')=\pm(0,\pi/2,\pi/4,3\pi/4)+C.
 \e{3.4b}

    Recall that we define both $\theta$ and $\phi$ modulo $2\pi$,
and therefore each measurement direction corresponds to two sets of
$(\theta, \phi)$ [see Eq.\ \rqn{2.7}]. Consequently, the optimal
configurations described by Eqs.\ (\ref{3.26a})--(\ref{3.4b}) can be
also described in several equivalent forms by applying the
transformation \rqn{2.7} to some of the four measurement directions.

    As follows from Eq.\ \rqn{3.26a}, the only odd states for which the
optimal vertical configurations lie in the same plane are the Bell
states $|\Psi_-\rangle$ (corresponding to $\alpha=\pi$) and
$|\Psi_+\rangle$ (corresponding to $\alpha=0$). The optimal vertical
configurations for the singlet state $|\Psi_-\rangle$ are given by
 \be
\phi_a=\phi_a'=\phi_b=\phi_b'=\phi_0
 \e{3.19}
and Eqs.\ (\ref{3.4a})--(\ref{3.4b}). To describe the optimal
vertical configurations for the state $|\Psi_+\rangle$, it is
natural to apply the equivalence \rqn{2.7} to the qubit $b$
measurement directions, so that the angles $\phi$ are still all
equal as in Eq.\ (\ref{3.19}), while the angles $\theta$ are given
by Eqs.\ (\ref{3.4a})--(\ref{3.4b}) with flipped signs for the qubit
$b$, i.e.\  $\theta_b\rightarrow -\theta_b$ and
$\theta_b'\rightarrow -\theta_b'$.

 Now let us consider the optimal detector configurations in the horizontal
($xy$) plane:
 \be
\theta_a=\theta_a'=\theta_b=\theta_b'=\pi/2.
 \e{3.10}
All configurations for $S=2\sqrt{2}$ can be obtained from the
standard configuration (\ref{conf-st-1})--\rqn{conf-st-2} by
rotating it into the $xy$ plane (so that the angles $\theta$ are
essentially replaced by the angles $\phi$), then applying an
arbitrary rotation within $xy$ plane and possibly the mirror
reflection, and finally applying the transformation \rqn{g4} with
the state-dependent parameter $\alpha$, so that
 \be
(\phi_a,\phi_a',\phi_b+\alpha,\phi_b'+\alpha)=
\pm(0,\pi/2,\pi/4,3\pi/4)+C
 \e{3.6a}
 with arbitrary $C$ (the signs $\pm$ correspond again to the possibility of the mirror reflection).

    Similarly, all horizontal configurations corresponding to $S=-2\sqrt{2}$
for the odd states are described by the angles
 \be
(\phi_a,\phi_a',\phi_b+\alpha,\phi_b'+\alpha)=
\pm(0,\pi/2,-3\pi/4,-\pi/4)+C.
 \e{3.6b}

    Notice that the application of the equivalence \rqn{2.7} to all four
 measurement directions changes $\pi/2$ into $-\pi/2$ in Eq.\
 \rqn{3.10}, while Eqs.\ \rqn{3.6a} and \rqn{3.6b} do not change,
 since the corresponding $\pi$-shift of angles $\phi$ can be absorbed
by the arbitrary parameter $C$.

\section{Local measurement errors}
\label{IV}

 In this section we consider the effects of local (independent)
measurement errors on the BI violation.

\subsection{Error model}

The probabilities of the measurement results for a single qubit can
be written in the form
 \be
p_i^M=\sum_{m=0}^1F_{im}p_m={\rm Tr}(Q_i\rho_q),
 \e{3.3}
where $p_m$ are the probabilities which would be obtained by ideal
measurements, $F_{im}$ is the probability to find the qubit in the
state $|i\rangle$ when it is actually in the state $|m\rangle$, and
operator $Q_i=F_{i0}P_0+F_{i1}P_1$ contains the projector operators
$P_{0,1}$ [see Eq.\ \rqn{12}].
 The operators $Q_i$ satisfy the same condition as the POVM measurement
operators, \cite{nie00} namely, $Q_i$ are positive and $Q_0+Q_1=1$.
 The condition $p_0^M+p_1^M=1$ implies that $F_{0m}+F_{1m}=1$.
 Hence, the matrix $F$ has two independent parameters, which can be chosen
as the measurement fidelities $F_0 \equiv F_{00}$ and $F_1\equiv
F_{11}$ for the states $|0\rangle$ and $|1\rangle$, so that
    \begin{equation}
p_0^M = F_0\tilde{\rho}_{00}+(1-F_1)\tilde{\rho}_{11}, \,\,\,
    p_1^M =(1-F_0)\tilde{\rho}_{00}+F_1\tilde{\rho}_{11}
    \end{equation}
(here $\tilde{\rho}_{ij}$ are the components of the one-qubit
density matrix after the tomographic rotation and $0\leq F_{0,1}\leq
1$). It can be always assumed that $F_0+F_1\geq 1$, since in the
opposite case the measurement results can be simply renamed:
$0\leftrightarrow 1$; as a consequence, $\max (F_0, F_1)\geq 1/2$.

Using the assumption that measurement errors for each qubit can be
considered independently of the errors for the other qubit, the
measured probabilities for a qubit pair can be written in the form
\cite{cla78}
 \be
p_{ij}^M=\sum_{m,n=0}^1F^a_{im}F^b_{jn}p_{mn} ={\rm Tr}(Q_i^a
Q_j^b\rho),
 \e{3.1}
where $F^k_{im}$ is the matrix $F_{im}$ for qubit $k$ and
$Q_i^k=F_{i0}^kP_0^k+F_{i1}^kP_1^k$ [see Eq.\ \rqn{13}].

    In this section we will discuss the condition for the BI violation as a
function of measurement fidelities $F_0^k$ and $F_1^k$. Sometimes we
will limit the analysis to the case of equal measurement fidelities
for both qubits,
 \be
F_i^a=F_i^b=F_i,
 \e{4.11}
however, the case of different measurement fidelities for the two
qubits is also of interest. (Different fidelities are especially of
interest when the qubits have different physical implementations.
For instance, in the case of an atom-photon qubit pair \cite{moe04}
the detection efficiency for the atom is nearly 100\%, whereas the
photon-detector efficiency is significantly less than 100\%.) Notice
that for the Josephson phase qubits the trade-off between the
fidelities $F_0^k$ and $F_1^k$ can be controlled in the experiment
\cite{mcd05,coo04} for each qubit individually by changing the
measurement pulse strength.

Several special cases of our error model have been previously
discussed in the literature, starting with the CHSH
paper.\cite{CHSH} For example, in the problem of the detector
loophole \cite{pea70,gar87} the CH inequality with $F_0^a=F_0^b=1$
is often considered; then $F_1^k$ is called the detector efficiency;
both the cases $F_1^a=F_1^b$ (Refs. \onlinecite{ebe93,lar01}) and
$F_1^b\ne F_1^a$ (Ref.\ \onlinecite{cab07}) have been considered.
Let us also mention the effect of nonidealities on the BI violation
considered for the experiments on two-photon interference.
\cite{kwi93,zuk93}
 The situations of Refs.\ \onlinecite{kwi93,zuk93} formally
 correspond  to the special case of our model with
 \be
F_0^a=F_1^a=F_a,\ F_0^b=F_1^b=F_b.
 \e{4.12}
Then the product $(2F_a-1)(2F_b-1)$ equals either the visibility
\cite{kwi93} or the product of the visibility and the square of the
signal acceptance probability. \cite{zuk93}

\subsection{General relations for $S$}
 \label{IVB}

The Bell operator (\ref{2.2}) can be generalized to the case of
measurement errors.
 Inserting Eq.\ \rqn{3.1} into Eq.\ \rqn{5} yields
$E(\vec{a},\vec{b})={\rm Tr}(\tilde{A}\tilde{B}\rho)$, where
 \bes{4.2}
 \bea
&\tilde{A}&=Q_1^a-Q_0^a =F_1^a-F_0^a+(F_0^a+F_1^a-1)\,
\vec{a}\cdot\vec{\sigma}_a, \qquad \,\,
 \label{4.2a} \\
&\tilde{B}&=F_1^b-F_0^b+(F_0^b+F_1^b-1)\,
\vec{b}\cdot\vec{\sigma}_b.
 \ea{4.2b}
 \ese
Therefore $S$ can be expressed as
 \be
S={\rm Tr}(\tilde{\cal B}\rho)
 \e{4.19}
via the modified Bell operator
 \be
\tilde{\cal B}=\tilde{A}\tilde{B}-\tilde{A}\tilde{B}'+
\tilde{A}'\tilde{B}+\tilde{A}'\tilde{B}',
 \e{4.3}
where $\tilde{A}'$ and $\tilde{B}'$ are obtained from $\tilde{A}$
and $\tilde{B}$ by replacing $\vec{a}$ and $\vec{b}$ with $\vec{a}'$
and $\vec{b}'$, respectively. Notice that $\tilde{\cal B}$ is a
Hermitian operator and therefore in some cases it is useful to think
about the measurement of $S$ as a measurement of a physical quantity
corresponding to the operator $\tilde{\cal B}$ (even though this
analogy works only for averages).

    It is rather trivial to show\cite{rem-bound} that in the presence
of local measurement errors the Cirel'son's inequality $|S|
\le2\sqrt{2}$ remains valid (this fact can be proven \cite{die02}
for any POVM-type measurement). Moreover, a stricter inequality for
$|S|$ [see Eq.\ (\ref{4.9}) below] can be obtained, using the method
similar to that of Ref. \onlinecite{die02}. We will prove this
inequality for all pure two-qubit states,
$\rho=|\psi\rangle\langle\psi|$, which automatically means that it
is also valid for any mixed state $\rho$. Using notation $\langle
O\rangle=\mbox{Tr}(O\rho )= \langle\psi|O|\psi\rangle$ for any
operator $O$, we start with the obvious relation $|S|
=|\langle\tilde{A}(\tilde{B}-\tilde{B}')\rangle
+\langle\tilde{A}'(\tilde{B}+\tilde{B}')\rangle|
\le|\langle\tilde{A}(\tilde{B}-\tilde{B}')\rangle|
+|\langle\tilde{A}'(\tilde{B}+\tilde{B}')\rangle|$.
    The next step is to apply the general inequality
$|\langle O_1O_2\rangle|^2 \le\langle O_1O_1^\dagger\rangle\langle
O_2^\dagger O_2\rangle$ to both terms in the sum (this inequality is
the direct consequence of the Cauchy-Schwartz inequality
$|\langle\psi_1|\psi_2\rangle|^2\le\langle
\psi_1|\psi_1\rangle\langle\psi_2|\psi_2\rangle$ for the vectors
$|\psi_1\rangle=O_1^\dagger|\psi\rangle$ and
$|\psi_2\rangle=O_2|\psi\rangle$). In this way we obtain
 \be
|S| \le\sqrt{\langle\tilde{A}^2\rangle
\langle(\tilde{B}-\tilde{B}')^2\rangle}
+\sqrt{\langle\tilde{A}'\,^2\rangle
\langle(\tilde{B}+\tilde{B}')^2\rangle}
 \e{4.4}
(notice that operators $\tilde{A}$, $\tilde{A}'$, $\tilde{B}$, and
$\tilde{B}'$ are Hermitian).
   As the next step, we notice that the eigenvalues of $\tilde{A}$
(as well as eigenvalues of  $\tilde{A}'$) are $2F_1^a-1$ and
$1-2F_0^a$, which follows from Eq.\ \rqn{4.2} and the fact that the
eigenvalues of $\vec{a}\cdot\vec{\sigma}_a$ are $\pm 1$. Therefore,
$\langle\tilde{A}^2\rangle \le (2F_{\rm max}^a -1)^2$ and
$\langle\tilde{A}'\,^2\rangle\le (2F_{\rm max}^a -1)^2$, where
$F_{\rm max}^k=\max(F_0^k,F_1^k)$; and so from Eq.\ (\ref{4.4}) we
obtain $|S|\le (2F_{\rm max}^a-1)
\left[\sqrt{\langle(\tilde{B}-\tilde{B}')^2\rangle}
+\sqrt{\langle(\tilde{B}+\tilde{B}')^2\rangle}\right]$. Next, since
$\sqrt{x_1}+\sqrt{x_2}\le\sqrt{2(x_1+x_2)}$ for any positive numbers
$x_1$ and $x_2$, and using the relation
    $\langle(\tilde{B}-\tilde{B}')^2\rangle
+\langle(\tilde{B}+\tilde{B}')^2\rangle = 2
\langle\tilde{B}^2+\tilde{B}'\,^2\rangle$, we obtain the inequality
$|S|\le 2 (2F_{\rm max}^a-1)
\sqrt{\langle\tilde{B}^2+\tilde{B}'\,^2\rangle}$.
   Finally, using the relations $\langle\tilde{B}^2\rangle \le (2F_{\rm max}^b -1)^2$ and
$\langle\tilde{B}'\,^2\rangle\le (2F_{\rm max}^b -1)^2$, derived in
a similar way as above, we obtain the upper bound
 \be
|S|\le2\sqrt{2}(2F_{\rm max}^a-1)(2F_{\rm max}^b-1).
 \e{4.9}
This upper bound is generally not exact and can be reached only in
the case when the errors are symmetric in both qubits [Eq.\
\rqn{4.12}], leading to Eq.\ \rqn{4.14} below. While the bound
\rqn{4.9} depends only on the largest measurement fidelity for each
qubit, our numerical results show that the exact bounds $S_{\pm}$
shrink monotonously with the decrease of all fidelities, if the
errors are small enough to allow the BI violation (see below).

A useful expression for $S$ can be obtained from Eqs.\
\rqn{4.2}--\rqn{4.19} by separating the terms for the ideal case:
 \be
S=2\xi_-^a\xi_-^b+2\xi_+^a\xi_-^b\vec{a}'\cdot\vec{s}_a
+2\xi_-^a\xi_+^b\vec{b}\cdot\vec{s}_b+\xi_+^a\xi_+^bS_0,
 \e{4.1}
where $\xi_+^k=F_0^k+F_1^k-1,\ \xi_-^k=F_1^k-F_0^k$, $S_0$ is the
value of $S$ in the absence of errors [Eq.\ \rqn{3.5}], and
$\vec{s}_k$ is the Bloch vector characterizing the reduced density
matrix for the qubit $k$, i.e.\ $\rho_k=\mbox{Tr}_{k'\ne
k}\rho=(I+\vec{s}_k\cdot\vec{\sigma})/2$.
    Notice that $\vec{s}_a=\vec{s}_b=0$ for a maximally entangled
state and therefore the second and third terms in Eq.\ \rqn{4.1} may
increase $|S|$ for nonmaximally entangled optimal states.
 That is why in the presence of errors the states maximizing and
minimizing $S$ are usually nonmaximally entangled \cite{ebe93} (see
below).

Notice that in the presence of errors, $S$ still preserves the
invariance with respect to the local transformations of qubits and
simultaneous rotation of measurement directions described by Eqs.\
\rqn{3.16}. The symmetry described by Eq.\ \rqn{3.25} (sign flip of
$S$ for the reversal of one-qubit measurement directions) is no
longer valid; however, it can be easily modified by adding
simultaneous interchange $F_0\leftrightarrow F_1$ of one-qubit
fidelities:
      \bes{trSF}
    \begin{eqnarray}
&& S \rightarrow -S \,\,\,\, \mbox{if} \,\,\,\, \vec{a}\rightarrow
-\vec{a},\,
  \vec{a}'\rightarrow -\vec{a}', \, F_0^a\leftrightarrow F_1^a ;
  \qquad
    \label{trSF-a} \\
&& S \rightarrow -S \,\,\,\, \mbox{if} \,\,\,\, \vec{b}\rightarrow
-\vec{b},\,
  \vec{b}'\rightarrow -\vec{b}', \, F_0^b\leftrightarrow F_1^b.
  \ea{trSF-b}
\ese
   Obviously, $S$ does not change if both transformations
(\ref{trSF}) are made simultaneously. As a consequence, the maximum
$S_+$ and minimum $S_-$ (optimized over the qubit states and over
measurement directions) are invariant with respect to simultaneous
interchange of measurement fidelities
 \be
F_0^a\leftrightarrow F_1^a,\ \ F_0^b\leftrightarrow F_1^b ,
 \e{4.8}
while the extremum values change as $S_+\rightarrow -S_-$,
$S_-\rightarrow -S_+$ if only one-qubit fidelity interchange
($F_0^a\leftrightarrow F_1^a$ or $F_0^b\leftrightarrow F_1^b$) is
made. (The corresponding optimal states obviously do not change.)

In the presence of measurement errors the magnitudes of the maximum
and minimum of $S$ generally differ, $S_+\ne |S_-|$. However, as
follows from the latter symmetry, $S_+ = |S_-|$ (as in the ideal
case) if the two measurement fidelities are symmetric (equal) at
least for one qubit:
 \be
F_0^a=F_1^a\ \,\,\, \mbox{or} \,\,\, F_0^b=F_1^b .
 \e{4.23}

    If the fidelities are symmetric for both qubits [the situation
described by Eq.\ \rqn{4.12}], then the expression for $S$ given by
Eq.\ \rqn{4.1} becomes simple: $S=(2F_a-1)(2F_b-1)S_0$ and directly
related to the value $S_0$ without measurement errors. Then the
extremum values
 \be
S_\pm=\pm2\sqrt{2}(2F_a-1)(2F_b-1)
 \e{4.14}
are obviously achieved for any maximally entangled state under the
same conditions as in Sec.\ \ref{III}.
 Correspondingly, the requirement for the fidelities for
a violation of the BI is \cite{kwi93,zuk93}
 \be
(2F_a-1)(2F_b-1)>2^{-1/2}\approx0.707.
 \e{4.5}
Notice that when the measurement fidelities for the both qubits are
the same, $F_a=F_b=F$, Eq.\ \rqn{4.5} reduces to the threshold
fidelity
 \be
F>0.5+2^{-5/4}\approx0.920 ,
 \e{4.20}
while if the measurement for one of the qubits is ideal (for
example, $F_a=1$), then the BI violation requires
 \be
F_b>0.5+2^{-3/2}\approx0.854.
 \e{4.10}

\subsection{Analytical results for maximally entangled states}
\label{IVC}

Let us first analyze the extremum values of $S$ for the class of
maximally entangled states (\ref{3.18}).
 Since in this case
$\vec{s}_a=\vec{s}_b=0$, we obtain from Eq.\ \rqn{4.1} that for
maximally entangled states
 \be
S=2(F_1^a-F_0^a)(F_1^b-F_0^b) +(F_1^a+F_0^a-1)(F_1^b+F_0^b-1)S_0
 \e{23}
is directly related to the corresponding quantity $S_0$ in the
absence of errors.
 Therefore the extremum values of $S$ for maximally entangles states are
 \be
S_\pm=2(F_1^a-F_0^a)(F_1^b-F_0^b)
\pm2\sqrt{2}(F_1^a+F_0^a-1)(F_1^b+F_0^b-1),
 \e{25}
and they are achieved under the same conditions as discussed in
Sec.\ \ref{III}.

  When the asymmetry of measurement fidelities is similar for both
qubits ($F_1^a>F_0^a$ and $F_1^b>F_0^b$ or both inequalities with
``$<$'' sign), the first term in Eq.\ \rqn{25} is positive, and
therefore the BI $|S|\leq 2$ can be stronger violated for positive
$S$ than for negative $S$. Similarly, if the asymmetries are
opposite (for example, $F_1^a>F_0^a$ and $F_1^b<F_0^b$), then it is
easier to violate the BI for negative $S$. If fidelities are
symmetric at least for one qubit [Eq.\ \rqn{4.23}], then the first
term in Eq.\ \rqn{25} vanishes, and therefore $S_-=-S_+$, as
discussed in the previous subsection.

    If the fidelities are the same for both qubits,
$F_0^a=F_0^b=F_0$ and $F_1^a=F_1^b=F_1$, then the positive $S$ is
preferable and
 \be
S_+=2(F_1-F_0)^2+2\sqrt{2}(F_1+F_0-1)^2
 \e{26}
(see the dashed lines in Fig.\ \ref{f1}).
 This value of $S_+$ reaches
the maximum $2\sqrt{2}$ when both errors vanish ($F_0=F_1=1$) and
decreases with the decrease of each fidelity in the interesting
region $S_+>2$ (more accurately, as long as $S_+>4-2\sqrt{2}\approx
1.17$).
 The condition for the BI violation in this case is
 \be
(F_1-F_0)^2+\sqrt{2}(F_1+F_0-1)^2>1.
 \e{27}
This threshold of the BI violation on the $F_0$-$F_1$ plane is shown
by the lowest dashed line in Fig.\ \ref{f1}. It is an arc of the
ellipse (corresponding to $S_+=2$), which is symmetric with respect
to the line $F_0=F_1$ and is centered at $F_0=F_1=0.5$. However, as
seen from Fig.\ \ref{f1}, this threshold looks quite close to a
straight line on the $F_0$-$F_1$ plane.
    Notice that in the case $F_0=1$ the threshold \rqn{27} reduces
to the most well-known condition \cite{CHSH,gar87}
 \be
F_1>2\sqrt{2}-2\approx0.828\, ,
 \e{4.7}
while in the case of symmetric error, $F_0=F_1=F$, we recover Eq.\
\rqn{4.20}.

\begin{figure}[htb]
\includegraphics[width=7.0cm]{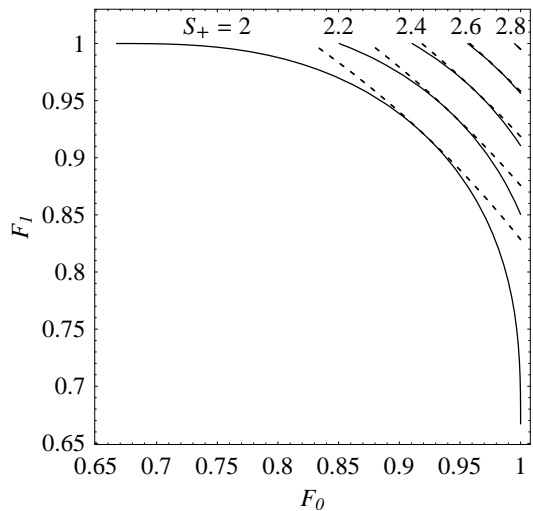}
\caption{Solid lines: contour plot of $S_+$ (the maximum quantum
value of $S$) versus the measurement fidelities $F_0$ and $F_1$
(assumed equal for both qubits), optimized over all two-qubit
states. The dashed lines show the result of $S_+$ maximization over
the maximally-entangled states only [Eq.\ \protect\rqn{26}]. The
Bell (CHSH) inequality can be violated when $S_+>2$. }
 \label{f1}\end{figure}

\subsection{Numerical results}

To optimize the CHSH inequality violation in presence of the
measurement errors over all two-qubit states, including
non-maximally entangled states, we have used numerical calculations.
The analysis has been performed in two different ways (with
coinciding results).
   First, we have searched for the maximum violation by finding
extrema of the eigenvalues of the modified Bell operator
$\tilde{\cal B}$ defined by Eq.\ (\ref{4.3}). Since $\tilde{\cal B}$
is a Hermitian operator and $S={\mbox Tr} (\tilde{\cal B}\rho )$,
for fixed measurement directions
$(\vec{a},\vec{a}',\vec{b},\vec{b}')$
 the maximum and minimum eigenvalues of $\tilde{\cal B}$ are
equal to the maximum and minimum values of $S$, optimized over the
two-qubit states. Therefore, optimization of the eigenvalues over
the measurement directions gives the extrema of $S$. Similar method
has been previously used \cite{ebe93} for the case of identical
local errors \rqn{4.11} with $F_0=1$, while we apply this method to
our more general error model.

The numerical maximization (minimization) of the largest (smallest)
eigenvalue of the Bell operator $\tilde{\cal B}$ has been performed
using the software package Mathematica. The full optimization should
be over all four measurement directions
($\vec{a},\vec{a}',\vec{b},\vec{b}')$, described by 8 angles total.
However, because of the invariance of $S$ under local
transformations, it is sufficient to optimize $\tilde{\cal B}$ over
only two angles: the angle between $\vec{a}$ and $\vec{a}'$ and the
angle between $\vec{b}$ and $\vec{b}'$, while the other angles are
kept fixed.

   Our numerical results show that in general the optimal
values of these two angles are different from each other; however,
for equal fidelity matrices [Eq.\ \rqn{4.11}] these angles are
equal, so that $\vec{a}\cdot \vec{a}'=\vec{b}\cdot \vec{b}'$.
 This result has been obtained previously \cite{ebe93} for the
special case $F_0=1$.
 Our numerical results also show that $S_+>|S_-|$  for positive
values of the product $(F_1^a-F_0^a)(F_1^b-F_0^b)$ and $S_+<|S_-|$
when this product is negative, similar to the result for the
maximally entangled states [see discussion after Eq.\ \rqn{25}].

    We have checked that the numerical results for $S_+$ and $|S_-|$
obtained via optimization of the eigenvalues of the Bell operator
$\tilde{\cal B}$ coincide with the results (see Fig.\ \ref{f1})
obtained by our second numerical method based on the direct
optimization of $S$. The second method happened to be more efficient
numerically; as another advantage, it provides the optimal
measurement directions together with optimal values $S_+$ and $S_-$,
while the Bell-operator method gives only $S_+$ and $S_-$.

    In principle, direct optimization of $S$ (for fixed measurement fidelities)
implies optimization over the two-qubit density matrix and over 8
measurement directions. However, there is a simplification. It is
obviously sufficient to consider only pure states, since
probabilistic mixtures of pure states cannot extend the range of
$S$. Moreover, it is sufficient to consider only the states of the
form
 \be
|\Psi\rangle=\cos(\beta/2)|10\rangle+\sin(\beta/2)|01\rangle,
 \e{4.16}
since any pure two-qubit states can be reduced to this form by local
rotations of the qubits (which are equivalent to rotations of the
measurement directions); this fact is a direct consequence of the
Schmidt decomposition theorem.\cite{nie00} The angle $\beta$ can be
limited within the range $0\le\beta\le\pi$ because the coefficients
of the Schmidt decomposition are non-negative. This range can be
further reduced to $0\le\beta\le\pi /2$ since $\pi$-rotation of both
qubits about $x$-axis (or any horizontal axis) exchanges states
$|10\rangle \leftrightarrow |01\rangle$ and therefore corresponds to
the transformation $\beta \rightarrow \pi-\beta$.

    Our numerical optimization of $S$ within the class
of two-qubit states (\ref{4.16}) has shown that for non-zero
measurement errors (we considered $2/3\leq F_i^k\leq 1$) the optimal
measurement directions ($\vec{a},\vec{a}',\vec{b},\vec{b}')$ always
lie in the same vertical plane [such configuration is described by
Eq.\ \rqn{3.19}]. This vertical plane can be rotated by an arbitrary
angle about $z$-axis [such rotation is equivalent to an overall
phase factor in Eq.\ \rqn{4.16}]; therefore we can assume $\phi_0=0$
in Eq.\ \rqn{3.19}. Notice that for the state \rqn{4.16} the vectors
$\vec{s}_a$ and $\vec{s}_b$ in Eq.\ \rqn{4.1} are along the
$z$-axis, $\vec{s}_a=-\vec{s}_b=\vec{z}\cos\beta$. These vectors are
zero for the maximally entangled state ($\beta =\pi/2$), then the
vertical configuration is no longer preferential; however, the
maximally entangled state is optimal only when there are no
measurement errors.

    For the state (\ref{4.16}) and vertical configuration \rqn{3.19}
of the detector axes the expression for $S$ has the form
 \bea
&&S=2\xi_-^a\xi_-^b-\xi_+^a\xi_+^b(g-h\sin\beta)\nonumber\\
&&+2\cos\beta(\xi_+^a\xi_-^b\cos\theta_a'
-\xi_-^a\xi_+^b\cos\theta_b),
 \ea{4.17}
 where
 \bea
&g=&\cos\theta_a\cos\theta_b-\cos\theta_a\cos\theta_b'
+\cos\theta_a'\cos\theta_b\nonumber\\
&&+\cos\theta_a'\cos\theta_b',\nonumber\\
&h=&\sin\theta_a\sin\theta_b-\sin\theta_a\sin\theta_b'
+\sin\theta_a'\sin\theta_b\nonumber\\
&&+\sin\theta_a'\sin\theta_b'.
 \ea{4.24}
The numerical maximization and minimization of $S$ in this case
involves optimization over 5 parameters: $\beta$, $\theta_a$,
$\theta_a'$, $\theta_b$, and $\theta_b'$. Nevertheless, in our
calculations this procedure happened to be faster than optimization
over only two parameters in the method based on the Bell operator
eigenvalues (we used Mathematica in both methods).

   The solid lines in Fig.\ \ref{f1} show the contour plot of maximum
value $S_+$ on the plane $F_0$-$F_1$ for the case when the
measurement fidelities for two qubits are equal [Eq.\ \rqn{4.11}; in
this case $S_+\ge|S_-|$]. Notice that the line for $S_+=2$ ends at
the points\cite{ebe93} $F_0=1,\, F_1=2/3$ and $F_0=2/3,\, F_1=1$
(strictly speaking, this line corresponds to $S_+=2+0$ since $S=2$
can be easily realized without entanglement).
   The dashed lines, which correspond to the optimization
over the maximally entangled states only [Eq.\ \rqn{26}], coincide
with the solid lines at the points $F_0=F_1$, because in this case
the optimum is achieved at the maximally entangled states, as
follows from the discussion after Eq.\ (\ref{4.23}). When $F_0\neq
F_1$, the use of non-maximally entangled states gives a wider range
of measurement fidelities allowing the BI violation. However, as
seen from Fig.\  \ref{f1}, the difference between the solid and
dashed lines significantly shrinks with the increase of $S_+$, so
that there is practically no benefit of using non-maximally
entangled states for the BI violation stronger than $S_+>2.4$.
  Notice that the solid and dashed lines in Fig.\ \ref{f1} are symmetric about
the line $F_0=F_1$ since the interchange $F_0\leftrightarrow F_1$
does not change $S_+$, as was discussed after Eq.\ (\ref{trSF}).

Numerical calculations show that in the case of equal fidelity
matrices, Eq.\ \rqn{4.11}, the optimal detector configurations for a
nonmaximally entangled state \rqn{4.16} have a ``tilted-X'' shape:
$\vec{a}=-\vec{b}'$ and $\vec{a}'=-\vec{b}$.
 In this case the number of parameters to be optimized in \rqn{4.17}
reduced from 5 to 3, significantly speeding up the numerical
procedure.

    Notice that each optimal configuration within the class of
states \rqn{4.16} corresponds to a 6-dimensional manifold of optimal
configurations, obtained by simultaneous local rotations of the
measurement axes and the two-qubit state (see discussion in Sec.\
\ref{IIIB2}).

\section{Decoherence}
\label{VI}

    The detailed analysis of the effects of decoherence will be
presented elsewhere. \cite{write} In this sections we discuss only
some results of this analysis, and also discuss the combined effect
of local measurement errors and decoherence.

To study effects of decoherence we assume for simplicity that the
qubit rotations are infinitely fast.
 Thus, we assume that after a fast preparation of a two-qubit
state $\rho$ there is a decoherence during time $t$ resulting in the
state $\rho'$, which is followed by fast measurement of $\rho'$
(including tomographic rotations).
 Now $S$ is given by Eq.\ \rqn{3.5} (in the absence of errors) or
\rqn{4.19} and \rqn{4.1} (in the presence of errors) where $\rho$
should be substituted by $\rho'$.
 To obtain $\rho'$ we assume independent (local) decoherence of each
qubit due to zero-temperature environment, described by the
parameters $\gamma_k=\exp (-t/T_1^k)$ and $\lambda_k=\exp
(-t/T_2^k)$ (here $k=a,b$) where $T_1^k$ and $T_2^k$ are the usual
relaxation times for the qubit $k$ ($T_{2}^k\leq 2T_1^k$).

As the initial state we still assume the state of the form
\rqn{4.16} (even though in presence of decoherence this state
actually does not always provide \cite{write} the extrema of $S$).
It can be shown analytically \cite{write} that in the absence of
measurement errors the maximum violation of the BI for the state
\rqn{4.16} can be achieved when the detector axes lie in either a
horizontal [Eq.\ \rqn{3.10}] or vertical [Eq.\ \rqn{3.19}] plane
(any other detector configuration cannot give stronger violation).
    In the case of only population relaxation ($T_2^k=2T_1^k$) the
horizontal configuration is better, while in the case of only
$T_2$-effect ($T_1^k=\infty$) the vertical configuration is better.

When local measurement errors are considered together with
decoherence, the optimal detector configurations may be neither
vertical nor horizontal.
 To elucidate this fact, note that in Eq.\ \rqn{4.1} with
$\rho$ replaced by $\rho'$ the vectors $\vec{s}_a$ and $\vec{s}_b$
remain vertical in the presence of decoherence.
 As a result, when in the absence of errors the optimal
configuration is horizontal, measurement errors may make the optimal
detector axes to go out of the horizontal plane.
 Note, however, that for some parameter ranges the vertical and
horizontal configurations are still optimal.

 In numerical calculations we should optimize $S$ over
8 parameters: $\beta$ and 7 detector angles (one of the angles
$\phi$ can be fixed because of the invariance of $S$ under identical
rotations of the qubits around the $z$ axis). We have performed such
optimization for a few hundred parameter points, choosing the
measurement fidelities $F^k_i$ and decoherence parameters $\gamma_k$
and $\lambda_k$ randomly from the range (0.7, 1).
  For many (more than half of) parameter points the optimal
configuration was still found to be either vertical or (in much
smaller number of cases) horizontal. Even when the optimal
configuration was neither vertical nor horizontal, we found that
restricting optimization to only the vertical and horizontal
configurations gives a very good approximation of the extrema
$S_\pm$ (within 0.01 for all calculated parameter points). Such
restriction significantly speeds up the calculations, since we need
to optimize over only 5 parameters instead of 8 parameters.

       Assuming initial state \rqn{4.16} and replacing $\rho$ by
$\rho'$ in Eq.\ \rqn{4.19} we obtain
 \bea
&S=&2\xi_-^a\xi_-^b+\xi_+^a\xi_+^b\{[1-\gamma_a-\gamma_b-
(\gamma_a-\gamma_b)\cos\beta]g\nonumber\\
&&+\lambda_a\lambda_bh\sin\beta\}
+2\xi_+^a\xi_-^b(\gamma_a+\gamma_a\cos\beta-1)\cos\theta_a'
\nonumber\\
&& +2\xi_-^a\xi_+^b(\gamma_b-\gamma_b\cos\beta-1)\cos\theta_b,
 \ea{6.4}
when the detector axes are in a vertical plane [$g$ and $h$ are
defined in Eq. \rqn{4.24}], and
 \bea
&&S=2\xi_-^a\xi_-^b+\xi_+^a\xi_+^b\lambda_a\lambda_b\sin\beta
[\cos(\phi_a-\phi_b)
\nonumber\\
&&\hspace{0.3cm} -\cos(\phi_a-\phi_b')+\cos(\phi_a'-\phi_b)
+\cos(\phi_a'-\phi_b')] \qquad
 \ea{6.2}
for a horizontal detector configuration.

 To find the extrema of $S$
within the class of vertical configurations, Eq.\ (\ref{6.4}) should
be numerically optimized over the parameter $\beta$ and four angles
$\theta$. The optimization of $S$ within the class of horizontal
configurations is much simpler, because the term in the square
brackets in Eq.\ \rqn{6.2} can be optimized independently of
$\beta$. This optimization is exactly the same as in the ideal case
[see Eqs.\ \rqn{3.6a} and \rqn{3.6b} with $\alpha =0$], therefore
the term in the square brackets has extrema $\pm 2\sqrt{2}$, and
therefore the extrema of Eq.\ \rqn{6.2} are reached at $\beta=\pi/2$
(maximally entangled state), thus yielding a rather simple formula
 \be
S_\pm=2\, \xi_-^a\xi_-^b\pm2\sqrt{2}\,
\xi_+^a\xi_+^b\lambda_a\lambda_b,
 \e{6.3}
which depends only on the $T_2$-relaxation and measurement
fidelities.

\begin{figure}[htb]
\includegraphics[width=7.0cm]{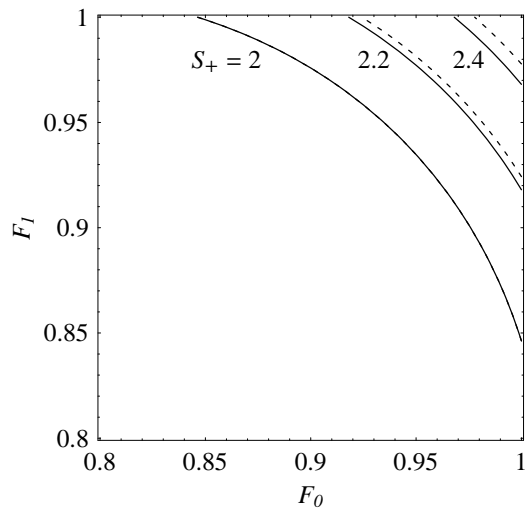}
\caption{Contour plot of the maximum value $S_+$ versus the
measurement fidelities $F_0$ and $F_1$ for the initial state
\rqn{4.16} in presence of decoherence with $\gamma_a=\gamma_b=0.96$
and $\lambda_a=\lambda_b=0.94$,
 in absence (solid lines) or presence (dashed lines) of the
symmetric crosstalk with $p_c=0.1$.
 Solid and dashed lines coincide for
$S_+=2$. }
 \label{f5}\end{figure}

For the numerical example shown by solid lines in Fig.\ \ref{f5} we
assume that decoherence is identical for the two qubits and choose
$\gamma_a=\gamma_b=0.96$ and $\lambda_a=\lambda_b=0.94$, that
corresponds to realistically good values for phase qubits:
$T_1\simeq 450$ ns, $T_2\simeq 300$ ns, and $t\simeq 20$ ns.
 We also assume identical errors for both qubits,  Eq.\ \rqn{4.11},
which implies $S_+\ge|S_-|$ (as in the absence of decoherence), so
in Fig.\ \ref{f5} we show the contour plot only for $S_+$.
 We have found that for these decoherence parameters the
vertical detector configuration is better than any other
configuration [assuming initial state \rqn{4.16}] for any
measurement fidelities in the analyzed range ($0.8\leq F_{0,1}\leq
1$). Notice that for the assumed decoherence parameters $S_+ =2.50$
in the absence of measurement errors, and the BI violation requires
$F>0.947$ (for $F_0=F_1=F$), that should be compared to the
threshold $F>0.920$ in absence of decoherence.

   Let us mention that the
error model previously discussed in relation to the BI violation in
the two-photon interference \cite{pav97} can be shown to be formally
equivalent to the special case of our model with pure dephasing
($T_1^k=\infty$) and identical errors with $F_0=1$. Then our
quantities $F_1$ and $\lambda_a\lambda_b$ correspond, respectively,
to the detector efficiency and visibility in Ref.
\onlinecite{pav97}.
 The case of pure dephasing in the absence of errors has been also
considered in connection with the BI violation in mesoscopic
conductors. \cite{sam03}

\section{Measurement crosstalk}
\label{VII}

The nonidealities discussed above are common for many types of
qubits.
  Now let us discuss a more specific type of error:
the measurement crosstalk for Josephson phase qubits.
\cite{mcd05,kof07} The crosstalk error originates from the fixed
(capacitive) coupling between the qubits, which is still on in the
process of measurement. The mechanism of the crosstalk is the
following.\cite{mcd05} If the measurement outcome for the phase
qubit $a$ is ``1'', then this qubit is physically switched to a
highly excited state (outside of the qubit Hilbert space), and its
dissipative oscillating evolution after the switching affects the
qubit $b$. As a result, the extra excitation of the qubit $b$ may
lead to its erroneous switching in the process of measurement, so
that instead of the measurement outcome ``1,0'' we may get ``1,1''
with some probability $p_c^a$. Similarly, because of the crosstalk
from the qubit $b$ to the qubit $a$, we may obtain the measurement
result ``1,1'' instead of ``0,1'' with some probability $p_c^b$. The
values of $p_c^a$ and $p_c^b$ significantly depend on the timing of
the measurement pulses applied to the qubits.\cite{mcd05} If the
qubit $a$ is measured few nanoseconds earlier than the qubit $b$,
then $p_c^a \gg p_c^b$; if the qubit $b$ is measured first, then
$p_c^a \ll p_c^b$. In the case when the measurement pulses are
practically simultaneous, the crosstalk probability becomes
significantly lower and $p_c^a \approx p_c^b$.

    Let us model the crosstalk in the following simple way. Even
though physically the crosstalk develops at the same time as the
measurement process and its description is quite nontrivial,
\cite{kof07} we will assume (for simplicity) that the crosstalk
effect happens after the ``actual'' measurement (characterized by
measurement fidelities as in Sections \ref{IV} and \ref{VI}), so
that the only effect of the crosstalk is the change of the outcome
``1,0'' into ``1,1'' with probability $p_c^a$ and the change of the
outcome ``0,1'' into ``1,1'' with probability $p_c^b$. Moreover, we
assume that the probabilities  $p_c^a$ and  $p_c^b$ do not depend on
the measurement axes ($\vec{a}$, $\vec{b}$, etc.).

 Notice that the measurement crosstalk obviously violates the fundamental
assumption of locality, on which the BI is based (so, strictly
speaking the BI approach is not applicable in such situation). In
this section we discuss the modification of the classical bound for
$S$, taking the crosstalk into account (this bound is now
model-dependent, in contrast to the usual BI), and we also analyze
the effect of the crosstalk on the quantum result for $S_\pm$.
Besides that, we discuss a simple modification of the experimental
procedure, which eliminates the effect of crosstalk by using only
the ``negative result'' outcomes.

\subsection{Modified Bell (CHSH) inequality}
 \label{VIA}

First, let us briefly review the derivation of the CHSH inequality,
presented in Ref. \onlinecite{asp02}.
 In a local realistic theory
 \be
S=\int s(\Lambda) F(\Lambda) \, d\Lambda ,
 \e{7.1}
where $F(\Lambda)$ is a distribution of the hidden variable
$\Lambda$ and
 \bea
&s (\Lambda )=&A(\Lambda,\vec{a})B(\Lambda,\vec{b})-
A(\Lambda,\vec{a})B(\Lambda,\vec{b}')\nonumber\\
&&+A(\Lambda,\vec{a}')B(\Lambda,\vec{b})+
A(\Lambda,\vec{a}')B(\Lambda,\vec{b}'). \quad
 \ea{7.2}
Here the measurement outcomes $A(\Lambda,\vec{a})$ and
$B(\Lambda,\vec{b})$ can take only the values $\pm1$, depending on
the hidden variable $\Lambda$ and the detector orientations for the
qubits $a$ and $b$. (Notice that in Sections \ref{III} and \ref{IV}
the notation $A$ and $B$ has been used for operators; now they are
classical quantities. Also notice that the outcome value $-1$ is
associated with the result ``0''.)
    It is easy to check that $s(\Lambda )=\pm 2$ under
the {\em locality assumption}: the result $A(\Lambda,\vec{a})$ does
not depend on the orientations $\vec{b}$ of the qubit $b$ and vice
versa; similarly, $F(\Lambda)$ does not depend on $\vec{a}$ and
$\vec{b}$. After integration (\ref{7.1}), this leads to the CHSH
inequality \rqn{3}.

   In our model the measurement crosstalk cannot change the positive product of outcomes
$AB=1$; however, it changes $AB=-1$ into $AB=1$ with the probability
$p_c^a (\Lambda )$ if $A=-B=1$ or with probability $p_c^b(\Lambda )$
if $A=-B=-1$. (The locality assumption is obviously violated, since
the value of $A$ now depends not only on $\Lambda$ and $\vec{a}$ but
also on $B$ and thus implicitly on $\vec{b}$; similarly for $B$.)
Notice that we assume that the crosstalk probabilities $p_c^a$ and
$p_c^b$ may in principle depend on $\Lambda$ (this is a slight
generalization of the more natural $\Lambda$-independent model for
$p_c^a$ and $p_c^b$).

    The random change from $AB=-1$ to $AB=1$ due to the crosstalk
leads to the modification of the CHSH inequality \rqn{3}.
  Here we mention only the main points of the derivation of
the modified CHSH inequality; the details are in the Appendix.
 We start with fixing $\Lambda$ and considering all possible changes of the
quantity $s (\Lambda )$ in Eq.\ \rqn{7.2} due to the crosstalk, thus
obtaining the new values of $s$ together with their probabilities
for each of 16 realizations of the vector ${\cal C}=(A,A',B,B')$.
 It is easy to see that due to the crosstalk $s (\Lambda )$ can get the values $\pm 4$,
which are outside of the limits $\pm 2$. Averaging the value $s
(\Lambda )$ over the crosstalk scenarios and then maximizing and
minimizing the result over 16 realizations of the vector ${\cal C}$,
we get
    \be
-2+4\min\{p_c^a (\Lambda ),p_c^b (\Lambda )\}\le \langle s (\Lambda
)\rangle \le2+2|p_c^a(\Lambda )-p_c^b(\Lambda)|.
 \e{6.6m}
Finally averaging this result over $\Lambda$, we obtain the modified
CHSH inequality:
 \be
-2+4\min\{p_c^a,p_c^b\}\le S\le2+2|p_c^a-p_c^b|,
 \e{6.6}
which is the main result of this subsection.

Let us consider two special cases.
 For a symmetric crosstalk,
$p_c^a=p_c^b=p_c$, the inequality \rqn{6.6} becomes
 \be
-2+4p_c\le S\le 2,
 \e{6.1}
while for a fully asymmetric crosstalk, $p_c^b=0$, it becomes
 \be
-2\le S\le2+2p_c^a.
 \e{6.7}
(similarly for $p_c^a=0$).

    It is interesting to notice that the inequality in the symmetric
case is more restrictive than the BI \rqn{3} (``easier'' negative
bound and no change of the positive bound). This is actually quite
expected because in the limiting case $p_c=1$ the crosstalk makes
all $AB$ products equal 1, so that $S=2$ always, as also follows
from  Eq.\ \rqn{6.1}. In contrast, for the fully asymmetric
crosstalk the inequality \rqn{6.7} is less restrictive than \rqn{3}
(``harder to violate'' positive bound and no change of the negative
bound). In the case of a finite crosstalk asymmetry, both the
positive and negative bounds change [Eq.\ (\ref{6.6})].

   Let us emphasize that in contrast to the derivation of the original
CHSH inequality, $s(\Lambda )$ may be significantly outside of the
range $(-2,2)$; it can get the values $s(\Lambda )=\pm 4$ for any
(symmetric or asymmetric) crosstalk. So the fact that the lower
bound for $S$ never decreases and the upper bound increases only
slightly for small crosstalk, is due to the statistical averaging of
the random increase and decrease of $s(\Lambda )$ due to the
crosstalk.

\subsection{Quantum calculation of $S$}

 In the quantum case the crosstalk changes the measurement
probabilities $p_{ij}\rightarrow p_{ij}^C$ as
 \bea
&&p_{00}^C=p_{00}, \, p_{10}^C=(1-p_c^a)p_{10}, \,
p_{01}^C=(1-p_c^b)p_{01},
\nonumber\\
&&p_{11}^C=p_{11}+p_c^ap_{10}+p_c^bp_{01}.
 \ea{7.5}
Then using the definitions \rqn{4} and \rqn{5} we obtain that the
measured value of $S$ becomes
 \be
S^C=2\tilde{p}_c+(1-\tilde{p}_c)S+2(p_c^a-p_c^b)[p_b(\vec{b})-p_a(\vec{a}')],
 \e{7.6}
where $\tilde{p}_c=(p_c^a+p_c^b)/2$, while $S$, $p_a(\vec{a}')$, and
$p_b(\vec{b})$ are the quantities obtained in the absence of
crosstalk [$p_a(\vec{a}')$ and $p_b(\vec{b})$ are defined after Eq.\
\rqn{7}].

 In a general case the maxima and minima $S^C_\pm$ of Eq.\ \rqn{7.6} can be
found numerically.
 To estimate the effect of the crosstalk, let us calculate Eq.\ \rqn{7.6} for
a maximally entangled state in the absence of errors and
decoherence.
 Then $p_a(\vec{a}')=p_b(\vec{b})=1/2$, and using $S_{\pm}=\pm 2\sqrt{2}$
we obtain
 \be
S^C_+=2\sqrt{2}-(2\sqrt{2}-2)\tilde{p}_c, \,\,\,
S^C_-=-2\sqrt{2}+(2\sqrt{2}+2)\tilde{p}_c.
 \e{6.10}
As we see, both extrema are affected by the crosstalk, making the
range narrower from both sides; however, the lower boundary is
affected much stronger than the upper boundary. Comparing Eq.\
(\ref{6.10}) with the modified CHSH inequality (\ref{6.6}), we see
that the lower bound shifts up for the quantum result always faster
than for the classical bound; therefore the gap between the quantum
and classical bounds always shrinks due to the crosstalk. The
classical-quantum gap at positive $S$ shrinks from both sides due to
the crosstalk.

 In the case of symmetric crosstalk, $p_c^a=p_c^b=p_c$, we can easily
consider non-maximally entangled states, measurement errors and
decoherence, since Eq.\ \rqn{7.6} in this case reduces to
$S^C=2p_c+(1-p_c)S$.
 Therefore, $S^C_\pm$ are simply related to the values $S_\pm$
 without crosstalk (but with measurement errors and decoherence):
 \be
S^C_\pm=2p_c+(1-p_c)S_\pm
 \e{6.9}
[similar dependence was used in Eq.\ (\ref{6.10})].  A violation of
the upper bound of the modified CHSH inequality \rqn{6.1} can be
observed when $S_+^C=2p_c+(1-p_c)S_+>2$, which yields $S_+>2$, while
a violation of the lower limit in Eq.\ \rqn{6.1} requires
$S_-^C=2p_c+(1-p_c)S_-<-2+4p_c$, which yields $S_-<-2$.
 Quite surprisingly, the symmetric crosstalk does not change the
conditions for the BI violation. (Of course, the violation of the
increased lower bound is not as convincing psychologically as the
violation on the increased upper bound.)

Let us discuss the combined effect of local errors, decoherence, and
symmetric crosstalk for the numerical example considered in Sec.\
\ref{VI}, assuming symmetric crosstalk with $p_c=0.1$.
 Now for the state \rqn{4.16} in the absence of local measurement
errors ($F_0=F_1=1$) we get $S_+^C=2.45$, which is slightly less
than the value $S_+=2.50$ obtained in the absence of the crosstalk.
The dependence of $S_+^C$ on the measurement fidelities $F_0$ and
$F_1$ in this case is illustrated by the dashed lines in Fig.\
\ref{f5}.
 In accordance with the above discussion, the solid line for $S_+=2$ and
the dashed line for $S_+^C=2$ coincide (the BI violation boundary is
not affected), while the comparison of the solid and dashed lines
for $S_+=2.2$ and $2.4$ shows that the crosstalk makes an
observation of a given BI-violating value of $S_+$ more difficult.

    Figure \ref{f6} shows a similar contour plot on the $F_0$-$F_1$
plane for the lowest quantum value $S_-^C$ assuming the same
parameters as in Fig.\ \ref{f5}. Comparing Figs.\ \ref{f5} and
\ref{f6} we see that it is more difficult to violate the lower
classical bound (even though it is now only $-1.6$) than the upper
classical bound of 2. This is because we assumed the same
measurement fidelities for both qubits, that generally shifts the
quantum result up [as in Eq.\ (\ref{25})]. Notice that in Fig.\
\ref{f6} the reference bound of $-2$ can still be violated, though
only for almost perfect fidelities.

\begin{figure}[htb]
\includegraphics[width=7.0cm]{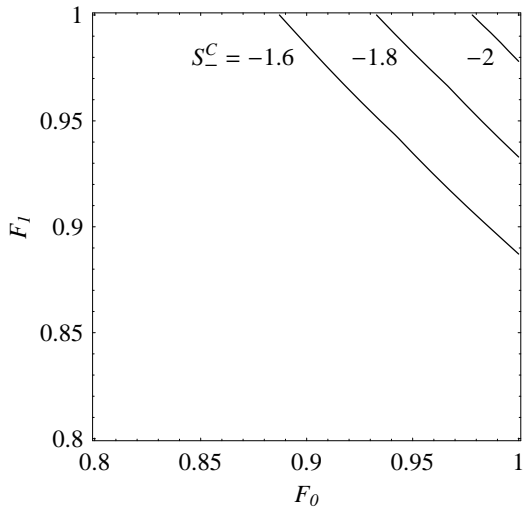}
\caption{Contour plot of the quantum minimum $S_-^C$ versus the
measurement fidelities $F_0$ and $F_1$ (same for both qubits)
optimized over the states \rqn{4.16}. We assume symmetric  crosstalk
with $p_c=0.1$ and decoherence parameters $\gamma_a=\gamma_b=0.96$
and $\lambda_a=\lambda_b=0.94$ (same as for Fig.\ \protect\ref{f5}).
The classical bound is shifted from $-2$ to $-1.6$ by the crosstalk.
}
 \label{f6}\end{figure}

\subsection{Elimination of the crosstalk effect}

    Slight modification of the CHSH inequality (\ref{3}) can make it
insensitive to the crosstalk. The main idea is to use only
experimental outcomes with the result ``0'', when a qubit does not
switch, and therefore the crosstalk does not occur. Such
``negative-result'' (``null-result'') experiments with the Josephson
phase qubits are very interesting from the quantum point of
view;\cite{negative-result,rus07} however, here we are interested
only in the classical consequence (or rather absence of it) for a
measurement with a null result.

    Instead of the inequality (\ref{3}) let us use the equivalent
inequality (\ref{6}), and let us change the definition of $T$ in
Eq.\ (\ref{7}) by interchanging the measurement outcomes ``1'' and
``0''. Then by symmetry the inequality (\ref{6}) is still valid, so
we get the classical bounds
    \begin{equation}
-1\leq \tilde{T}\leq 0
    \label{crt-1}    \end{equation}
 for
    \begin{eqnarray}
 &   \tilde{T}=p_{00}(\vec{a},\vec{b})-p_{00}(\vec{a},\vec{b'})+p_{00}(\vec{a'},\vec{b})
    +p_{00}(\vec{a'},\vec{b'})
    \nonumber \\
 &  -p_{0}(\vec{a'})-p_{0}(\vec{b}),
    \label{crt-2}\end{eqnarray}
where the probability $p_{0}(\vec{a'})$ is for measuring the qubit
$a$ only (without measuring the qubit $b$), while $p_{0}(\vec{b})$
is for measuring only the qubit $b$.

    With this simple modification, the CHSH inequality becomes
insensitive to the mechanism of the measurement
crosstalk\cite{mcd05,kof07} considered in this section. Notice,
however, that in performing experiment in this way it is still
important to check the absence of a direct crosstalk (due to the
measurement pulse itself). This can be done by applying a
measurement pulse to the well-detuned qubit $b$ (so that it cannot
switch) and checking that this does not affect switching
probabilities for the qubit $a$ (and similarly for $a$ interchanged
with $b$).

\section{Conclusion}
 \label{VIII}

    In this paper we have considered the conditions for the violation
of the Bell inequality in the CHSH form\cite{CHSH} for the entangled
pairs of solid-state qubits, when instead of the rotation of optical
polarizers (detectors) we have to rotate the states of two qubits
before the measurement, which itself is always performed in the
logical $z$-basis ($|0\rangle$,$|1\rangle$) for each qubit. While
most of our results are applicable to many types of qubits, we have
focused on the experiments with the Josephson phase qubits.
\cite{coo04} We have analyzed the BI violation for the ideal case as
well as in presence of various nonidealities, including local
measurement errors, local decoherence, and measurement crosstalk.

    In the ideal case the maximum violation of the BI ($S_\pm =\pm 2\sqrt{2}$,
while the classical bound is $|S|\leq 2$) can be realized for any
maximally entangled state. The optimal configuration of the
measurement directions in this case can be realized with three
degrees of freedom for each maximally-entangled state (the
measurement direction in our terminology actually refers to the
qubit rotation before the measurement). However, in presence of
nonidealities there is typically less freedom in choosing the
optimal configuration. For the ``odd'' two-qubit states involving
superpositions \rqn{4.16} of the states $|01\rangle$ and
$|10\rangle$, we have focused on the ``vertical'' measurement
configurations, for which all measurement directions
($\vec{a},\vec{a}',\vec{b},\vec{b}')$ are within the same vertical
plane of the Bloch sphere, and the ``horizontal'' configuration, for
which the four measurement axes are within the $x$-$y$ plane.

    The qubit measurement with finite local errors
(characterized by the fidelities $F_0$ and $F_1$ for each qubit)
shrinks the quantum range for $S$.
 We have found that for a maximally entangled
state the BI violation is still possible when the classical bounds
$\pm 2$ are exceeded by the extrema of the quantum result given by
Eq.\ (\ref{25}). In particular, when two qubits have the same
fidelities, the violation condition is given by Eq.\ (\ref{27}) and
shown by the lowest dashed line in Fig.\ \ref{f1}; it can be crudely
approximated by the condition $(F_1+F_0)/2>0.92$. A significantly
softer violation condition can be obtained when allowing the
two-qubit state to be non-maximally entangled; \cite{ebe93} this
condition is shown by the lowest solid line in Fig.\ \ref{f1}.
However, the trick of using a non-maximally-entangled state does not
help much when we need a BI violation with a significant margin (not
just barely); this can be seen by comparing the solid and dashed
lines in Fig.\ \ref{f1}.
   For non-maximally entangled odd two-qubit states \rqn{4.16} in presence
of local measurement errors, the vertical measurement configuration
is found to be preferable in comparison with other configurations.

    Analyzing the effect of local decoherence of the qubits for
the odd two-qubit states \rqn{4.16}, we have found that either
vertical or horizontal configuration of the measurement directions
is optimal, depending on the parameters. In particular, in the case
of population (energy) relaxation in $z$-basis, the horizontal
configuration is optimal, while for pure dephasing the vertical
configuration is optimal. In presence of both the decoherence and
local measurement errors, the optimal configuration can be neither
horizontal nor vertical; however, restricting optimization to only
these two classes of configurations gives a very good approximation
of the extrema $S_\pm$. Obviously, both the decoherence and the
measurement errors make the observation of the BI violation more
difficult.

    We have also analyzed the effect of the measurement crosstalk
\cite{mcd05} which plays an important role in measurement of the
capacitively-coupled phase qubits. Since the crosstalk is a
mechanism of classical communication between the qubits, strictly
speaking the BI is inapplicable. However, for a particular model of
the crosstalk it is possible to derive a modified CHSH inequality
[see Eq.\ (\ref{6.6})]. In particular, we have found that the
symmetric crosstalk does not change the upper classical bound but
increases the lower classical bound. The crosstalk also affects the
quantum bounds, which are given by Eq.\ (\ref{6.10}) for the
maximally entangled state in the otherwise ideal case with arbitrary
crosstalk and by Eq.\ (\ref{6.9}) for an arbitrary case but assuming
a symmetric crosstalk. Quite unexpectedly, the symmetric crosstalk
does not change the threshold condition for the observation of the
BI violation. However, the crosstalk always reduces the gap between
the classical and quantum bounds and makes an observation of the BI
violation with a finite margin more difficult. It is important to
mention that the detrimental effect of the crosstalk can be
eliminated by a slight change of the CHSH inequality (by using only
negative-result outcomes), which makes it insensitive to the
crosstalk [see Eqs.\ (\ref{crt-1}) and (\ref{crt-2})].

    We have performed the numerical simulations with the parameters similar
to the experimental values for the best present-day experiments with
the Josephson phase qubits. \cite{ans07,ste06} Our results (see
Fig.\ \ref{f5}) show the possibility of the CHSH inequality
violation with a significant margin even without further improvement
of the phase qubit technology.

\acknowledgments
    We thank Qin Zhang, John Martinis, Nadav Katz, and Markus
Ansmann for useful discussions.
 The work was supported by NSA and DTO under ARO grant
W911NF-04-1-0204.

\appendix*
\section{Derivation of the inequality \rqn{6.6}}

To derive the inequality \rqn{6.6m}, we fix $\Lambda$ and use the
abbreviated notation $s=s(\Lambda)$, $p_c^a=p_c^a(\Lambda)$,
$p_c^b=p_c^b(\Lambda)$, $A=A(\Lambda,\vec{a}),\
A'=A(\Lambda,\vec{a}')$ and similarly for $B$ and $B'$.
 Let us introduce the vectors ${\cal C}=(A,A',B,B')$ and
${\cal A}=(AB',AB,A'B,A'B')$, so that $s={\cal A}\cdot(-1,1,1,1)$.
 The vector ${\cal C}$ can assume 16 values, whereas without
the crosstalk ${\cal A}$ can take 8 values, since the number of
pluses or minuses in ${\cal A}$ is even.
 Each pair of ${\cal C}$ and $-{\cal C}$ yields one value of ${\cal
 A}$.
 Generally, the crosstalk effect differs for ${\cal C}$ and $-{\cal C}$, except
for the symmetric case, $p_c^a=p_c^b=p_c$, when crosstalk depends
only on ${\cal A}$.
 We start the analysis with the symmetric case and then consider the
general asymmetric case.

\subsection{Symmetric crosstalk}
 \label{A1}

 It is easy to check that without crosstalk $s=2$ or $-2$.
 To obtain the upper bound of the modified BI, we consider four
values of ${\cal A}$ corresponding to $s=2$. The crosstalk cannot
change ${\cal A}=(1,1,1,1)$, hence we discuss three other values:
$(-1,-1,1,1),\ (-1,1,-1,1)$, and $(-1,1,1,-1)$.
 For any of these vectors, the crosstalk makes $s$ to take the
values 0, 2, and 4 with the probabilities $p_cq_c,\ q_c^2+p_c^2$,
and $p_cq_c$, respectively, where $q_c=1-p_c$.
 Indeed, the change of any $-1$ to 1 in ${\cal A}$ occurs with the probability
$p_cq_c$, yielding $s=0$ for the change of the first $-1$ in ${\cal
A}$ and $s=4$ for the change of any other $-1$, whereas
$q_c^2+p_c^2$ is the probability of no change or change to ${\cal
A}=(1,1,1,1)$, both cases yielding $s=2$.
 Thus, though now $s$ can achieve the maximal mathematically possible
value 4, it is easy to see that the maximal value for the average of
$s$ over crosstalk is still $\langle s\rangle_{\rm max}=2$.

To obtain the lower limit of the inequality, we consider four values
of ${\cal A}$ corresponding to $s=-2$ in the absence of the
crosstalk.
 Consider first ${\cal A}=(-1,-1,-1,-1)$.
 This vector can be changed by the crosstalk to any of 16 possible
combinations of four pluses and minuses, yielding the values
$s=-4,-2,0,2,4$ with the probabilities $p_cq_c^3,\
q_c^4+3p_c^2q_c^2,\ 3p_cq_c^3+3p_c^3q_c,\ 3p_c^2q_c^2+p_c^4,\
p_c^3q_c$, respectively (note that the sum of the above
probabilities equals 1).
 The above probabilities can be easily obtained if one takes into
account that $s=-4$ results from the change of only the first $-1$
in ${\cal A}$, $s=-2$ occurs when either ${\cal A}$ have not changed
or the first and one of the last 3 components have changed, $s=0$
occurs when the crosstalk results in ${\cal A}$ with the first and
two other components equal to 1 or $-1$, $s=2$ occurs when two of
the last three components or all the components of ${\cal A}$ change
the sign, and $s=4$ results from the changes of all the last three
components in ${\cal A}$.
 As a result, in the case when only ${\cal A}=(-1,-1,-1,-1)$ is
 realized, we obtain
 \be
\langle s\rangle=-2+4p_c.
 \e{7.4}
 Finally, let us consider the values
${\cal A}=(1,-1,-1,1)$, $(1,1,-1,-1)$, and $(1,-1,1,-1)$.
 For any of these vectors the crosstalk makes $s$ to take the
values $-2$, 0, 2 with the probabilities $q_c^2,\ 2p_cq_c$, and
$p_c^2$, respectively.
 These probabilities follow from the fact that $s=-2$ when ${\cal
A}$ is not changed, $s=0$ results from the change of only one
component $-1$, and $s=2$ results from the change of the both
negative components.
 It is easy to check that for any of the three above vectors, we
again obtain Eq.\ \rqn{7.4}.

 Combining the results for the upper and lower bounds, we obtain the inequality
$-2+4p_c\le\langle s\rangle\le2$, the average of which over
$\Lambda$ yields the modified BI for the symmetric crosstalk given
by Eq.\ \rqn{6.1}.

\subsection{Asymmetric crosstalk}

As mentioned above, when $p_c^a\ne p_c^b$, the crosstalk yields
different results for ${\cal C}$ and $-{\cal C}$.
 However, recalling that the change
$AB=-1\rightarrow1$ occurs with the probability $p_c^a$ if $A=1$ and
$B=-1$ or with the probability $p_c^b$ if $A=-1$ and $B=1$, we can
obtain a symmetry relation for the probability of a change of $\cal
A$ with a given ${\cal C}$ to some ${\cal A}'$ as a function of
$p_c^a$ and $p_c^b$:
 \be
P_{{\cal A}\rightarrow{\cal A}'}(p_c^a,p_c^b,{\cal C}) =P_{{\cal
A}\rightarrow{\cal A}'}(p_c^b,p_c^a,-{\cal C}).
 \e{6.5}
To extend the results of Sec.\ \ref{A1} to the general case of
asymmetric crosstalk, we should reconsider all possible changes of
$\cal A$ discussed in Sec. \ref{A1} with the account of the two
vectors ${\cal C}$ and $-{\cal C}$ corresponding to each $\cal A$.
 As a result, the probability of each value of $s$ obtained above
is replaced by two probabilities, which differ from each other by
the change $p_c^a\leftrightarrow p_c^b$ [cf. Eq.\ \rqn{6.5}].

It happens that for most combinations ${\cal A}$ the probability
obtained in Sec. \ref{A1} is replaced by two probabilities just by
replacing $p_c$ with $p_c^a$ or with $p_c^b$.
 One of two exceptions is the case ${\cal A}=(-1,1,-1,1)$.
 Then for ${\cal C}=(-1,1,-1,1)$ the quantity $s=2$ changes
to $s=0,2,4$ with the probabilities $q_c^ap_c^b,\
q_c^aq_c^b+p_c^ap_c^b$, and $p_c^aq_c^b$, respectively, where
$q_c^k=1-p_c^k$.
 This yields $\langle s\rangle=2+2(p_c^a-p_c^b)$.
 The other value of $\langle s\rangle$ [corresponding to
${\cal C}=(1,-1,1,-1)$] is obtained from this formula by
substitution $p_c^a\leftrightarrow p_c^b$, so that $\langle
s\rangle=2+2(p_c^b-p_c^a)$.
 The other exception is ${\cal A}=(1,-1,1,-1)$.
 In this case we obtain $\langle s\rangle=-2+2(p_c^a+p_c^b)$ (for
both possible values of ${\cal C}$).
 Finally, by choosing the worst cases for the lower and upper bounds
for $\langle s\rangle $ and averaging the result over $\Lambda$, we
obtain Eq.\ \rqn{6.6}.


\begin{thebibliography}{99}
\bibitem{note4}
This parametrization is performed as follows.
 In view of Eq.\ \rqn{g2}, any maximizing detector configuration is
determined by $\vec{a}$ and $\vec{a}'$, and hence can be
equivalently described by a right-handed system of axes
$(\vec{X},\vec{Y},\vec{Z})$ such that $\vec{Z}=\vec{a},\
\vec{X}=\vec{a}',\ \vec{Y}=\vec{a}\times\vec{a}'$.
 This configuration can be obtained from the standard configuration
(\ref{conf-st-1})--(\ref{conf-st-2}) (with the corresponding frame
$\vec{X}_0=\vec{x},\ \vec{Y}_0=\vec{y},$ and $\vec{Z}_0=\vec{z}$) by
the axes rotation described by the Euler angles $\kappa_1$,
$\kappa_2$, and $\kappa_3$.
 This rotation is performed in three stages:
(i) the rotation around $\vec{Z}_0$ by $\kappa_1$, (ii) the rotation
around the new position of $\vec{Y}_0$ by $\kappa_2$, (iii) the
rotation around $\vec{Z}$ by $\kappa_3$.
\bibitem{note1}
These transformations are equivalent to the rotations described by
the Euler angles $(\kappa_1,\kappa_2,\kappa_3)=(\phi_0,C,0)$ and
$(\phi_0+\pi,-C,0)$ where $\phi_0$ and $C$ are arbitrary angles.
\bibitem{rem-bound} Our measurement error model for the qubit $k$
can be interpreted as the exchange of the measurement results
$0\leftrightarrow 1$ with probability $1-\max (F_0^k,F_1^k)$ and the
predetermined result with probability $|F_0^k-F_1^k|$ (result 0 if
$F_0^k>F_1^k$ or 1 if $F_0^k>F_1^k$), while the ``correct''
operation is realized with the remaining probability $\min (F_0^k,
F_1^k)$. Since the bound $|S|\leq 2\sqrt{2}$ obviously remains valid
in all the cases when one or both qubits operate in the mode of
result exchange or in the mode of predetermined result, and since
the erroneous operation modes are independent of the correct
results, the averaging over these modes (for each qubit) cannot
invalidate the bound $|S|\leq 2\sqrt{2}$.
\end{thebibliography}
\end{document}